\documentclass[prb,showpacs,superscriptaddress,amsmath,amssymb,twocolumn]{revtex4-1}

\usepackage{graphicx}
\usepackage[breaklinks=true,colorlinks=true,linkcolor=blue,urlcolor=blue,citecolor=blue]{hyperref}
\usepackage{amssymb}
\usepackage{amsmath}
\usepackage{mathtools}
\usepackage{graphicx}
\usepackage{tabularx}
\usepackage{multirow}	
\usepackage{ltxtable}
\usepackage{array}	
\usepackage{float}
\usepackage{epsfig}
\usepackage[caption=false]{subfig}
\usepackage{genmpage}
\usepackage{mathrsfs}
\usepackage{rotating}
\usepackage{latexsym}
\usepackage{fancyhdr} 	
\usepackage{enumerate}
\usepackage{setspace}
\usepackage{bm}
\usepackage{color}		
\usepackage{dcolumn}	
\usepackage{varwidth}
\usepackage{diagbox}
\usepackage{soul}		

\usepackage{tikz}
\usetikzlibrary{shapes.geometric, arrows}
\tikzstyle{startstop} = [rectangle, rounded corners, minimum width=3cm, minimum height=1cm,text centered, draw=black, fill=red!30, align=center]
\tikzstyle{io} = [trapezium, trapezium left angle=70, trapezium right angle=110, minimum width=3cm, minimum height=1cm, text centered, draw=black, fill=blue!30, align=center, inner sep=8pt]
\tikzstyle{process} = [rectangle, minimum width=1cm, minimum height=1cm, text centered, draw=black, fill=orange!30, align=center]
\tikzstyle{decision} = [diamond, minimum width=3cm, minimum height=1cm, text centered, draw=black, fill=green!30, align=center]
\tikzstyle{arrow} = [thick,->,>=stealth,]

\begin{document}

\title[]{Microscopy as a statistical, R\'enyi-Ulam, half-lie game: a new heuristic search strategy to accelerate imaging}

\author{Daniel W. Drumm}
\email{daniel.drumm@rmit.edu.au}
\affiliation{Australian Research Council Centre of Excellence for Nanoscale BioPhotonics, Physics, School of Science, RMIT University, Melbourne VIC 3000, Australia}

\author{Andrew D. Greentree}
\affiliation{Australian Research Council Centre of Excellence for Nanoscale BioPhotonics, Physics, School of Science, RMIT University, Melbourne VIC 3000, Australia}

\date{\today}

\begin{abstract}
Finding a fluorescent target in a biological environment is a common and pressing microscopy problem.  This task is formally analogous to the canonical search problem.  In ideal (noise-free, truthful) search problems, the well-known binary search is optimal.  The case of half-lies, where one of two responses to a search query may be deceptive, introduces a richer, R\'enyi-Ulam problem and is particularly relevant to practical microscopy.  We analyse microscopy in the contexts of R\'enyi-Ulam games and half-lies, developing a new family of heuristics.  We show the cost of insisting on verification by positive result in search algorithms; for the zero-half-lie case bisectioning with verification incurs a 50\% penalty in the average number of queries required.  The optimal partitioning of search spaces directly following verification in the presence of random half-lies is determined.  Trisectioning with verification is shown to be the most efficient heuristic of the family in a majority of cases.
\end{abstract}

\maketitle
\thispagestyle{fancy}


\section{Introduction}
\label{sec:intro}

Optical microscopy remains a key platform technology enabling detection, tracking, and sometimes quantification of biologically and medically relevant targets.  Standard and advanced microscopy approaches include confocal,\cite{Pawley06} and multiphoton\cite{Horton13} techniques, which deliver high-power light to classically small (diffraction-limited) spot volumes. The probe light excites either endogenous targets,\cite{Chalfie94} or introduced emitters (suitably functionalised to bind targets\cite{Reineck16,Reineck16a} within or between cells).

Standard approaches to confocal and multiphoton microscopy involve rastering the focal spot through a sample at constant scan rate; this can be treated as a finite dwell time on each pixel/voxel.  Firstly, this is to allow enough time for signal generation, and secondly, the dwell time is usually above a greater threshold value long enough that even dim regions of the sample exhibit low noise.  However, biological materials (and some introduced emitters) often exhibit photosensitive or even phototoxic responses\cite{Kochevar81}, and may photobleach over time\cite{Song95,Shaner08,Reineck16a}.  There is therefore a tension between resolving the image and minimally affecting the sample, and consequently a need to use the probe microscope light in the most efficient manner possible, extracting maximal information per photon used in both excitation and collection.

Whether using multiphoton or confocal techniques, a measurement illuminates a continuous sample region and is equivalent to asking a question of presence or absence of some target.  However, both confocal and multiphoton microscopes suffer considerable signal photon loss\cite{Pawley06}, due to their collection angles, detection point-spread functions, and sundry loss pathways.  Conversely, due to improvements in detector technology, false photon gain events (\textit{e.g.} dark counts) are now extremely rare.  Therefore the physical situation strongly resembles a half-lie scenario, where the presence of a particle may be subject to a lie (\textit{i.e.}, be undetected), but the absence of a particle is correctly reported -- albeit by an absence of signal.

Half-lies have been studied in the context of games where a Questioner attempts to guess which integer the Responder has in mind of those in some domain (\textit{e.g.}, [1,10$^{6}$]).  Only queries requiring yes/no answers are allowed; hence the response space is binary, and the category of problem is known as binary search\cite{Nowak08,Nowak09}.  The optimal classical method for the Questioner to accomplish this aim is to consecutively halve that domain; this technique is a special case of the splitting algorithm.\cite{Garey74}  For binary searches, $\lceil \log_{2}\left(m\right)\rceil$ queries suffice to locate one of $m$ integers.  The game becomes more interesting when the Responder occasionally lies.\cite{Renyi61,Ulam76}  These R\'enyi-Ulam games have been studied extensively, both mathematically,\cite{Pelc02,Ellis04,Ellis05,Ellis08,Cicalese13} and in various applications.\cite{Yao85,Ravikumar87,Lakshmanan91,Feige94,DeBonis97,Ngo00,Cicalese03,Mancini05,Karp07,Jedynak11,Corsi16}

As originally described, the R\'enyi game is of the combinatorial form ``Find a number $a\in\left\{1,2,...,m\right\}$'' using arbitrary queries with yes/no answers, where the Responder will lie a set number of times.\cite{Renyi61}  Ulam's game is similar, but specifies the number of times the Responder may lie (without enforcing such behaviour).  Variants since considered include asymmetric errors (including half-lies), asking only comparison-type questions, \textit{etc.} ([\onlinecite{Pelc02}] gives a comprehensive list).  Partially fidelitous solutions have been presented for a subset of these problems,\cite{Pelc89} though most studies to date have focussed upon finding optimal strategies where a minimum number of queries suffice in all cases to identify the target with 100\% fidelity.\cite{Rivest80,Ravikumar84,Ravikumar87,Pelc87,Lakshmanan91,Dhagat92,Muthukrishnan94,Ngo00,Cicalese00,Cicalese03,Cicalese03a,Cicalese04,Ellis04,Ellis05,Mancini05,Cicalese07,Ellis08,Xing16}

An interesting variant of R\'enyi-Ulam games arises when the lies are limited to only one response type, \textit{e.g.}, ``yes'', or ``present''.  Such constraints are termed ``half-lies'',\cite{Rivest80} and have been the subject of some enquiry,\cite{Lakshmanan91,Cicalese00} albeit for specific known numbers of such half-lies, or for unfettered choice of query.\cite{Cicalese13}  Another interesting subset of the games are those where the lies are no longer limited in number, but instead occur probabilistically\cite{Renyi61,Schalkwijk71,Pelc89} (otherwise known as the statistical variant).  We focus upon the intersection of these two sub-types.

Here, we connect microscopy and R\'enyi-Ulam games, and explore subtle differences from standard R\'enyi-Ulam games arising from the physics of confocal and multiphoton apparatus.  These differences preclude the usual approach of determining an optimal number of queries, instead forcing us to consider their average number across many trials.  We find a useful heuristic for accelerated searching under these conditions, which include the treatment of noise as an intrinsic characteristic of the measurement apparatus, rather than as a background distractor as assumed by signal detection theory\cite{Peterson54,Green66}.

This paper is organised as follows: in Sec. \ref{sec:game} we describe the equivalence between R\'enyi-Ulam games and microscopy, before developing a mathematical approach to selecting queries in Sec. \ref{sec:maths} and obtaining an information map.  We discuss potential uses of the map in various strategies in Sec. \ref{sec:search}, analysing their average behaviour across many trials, before concluding with Sec. \ref{sec:conc}.


\section{Microscopy as a R\'enyi-Ulam game} 
\label{sec:game}

\begin{figure}[tb!]
\centering
\includegraphics[angle=0, width=0.99\linewidth]{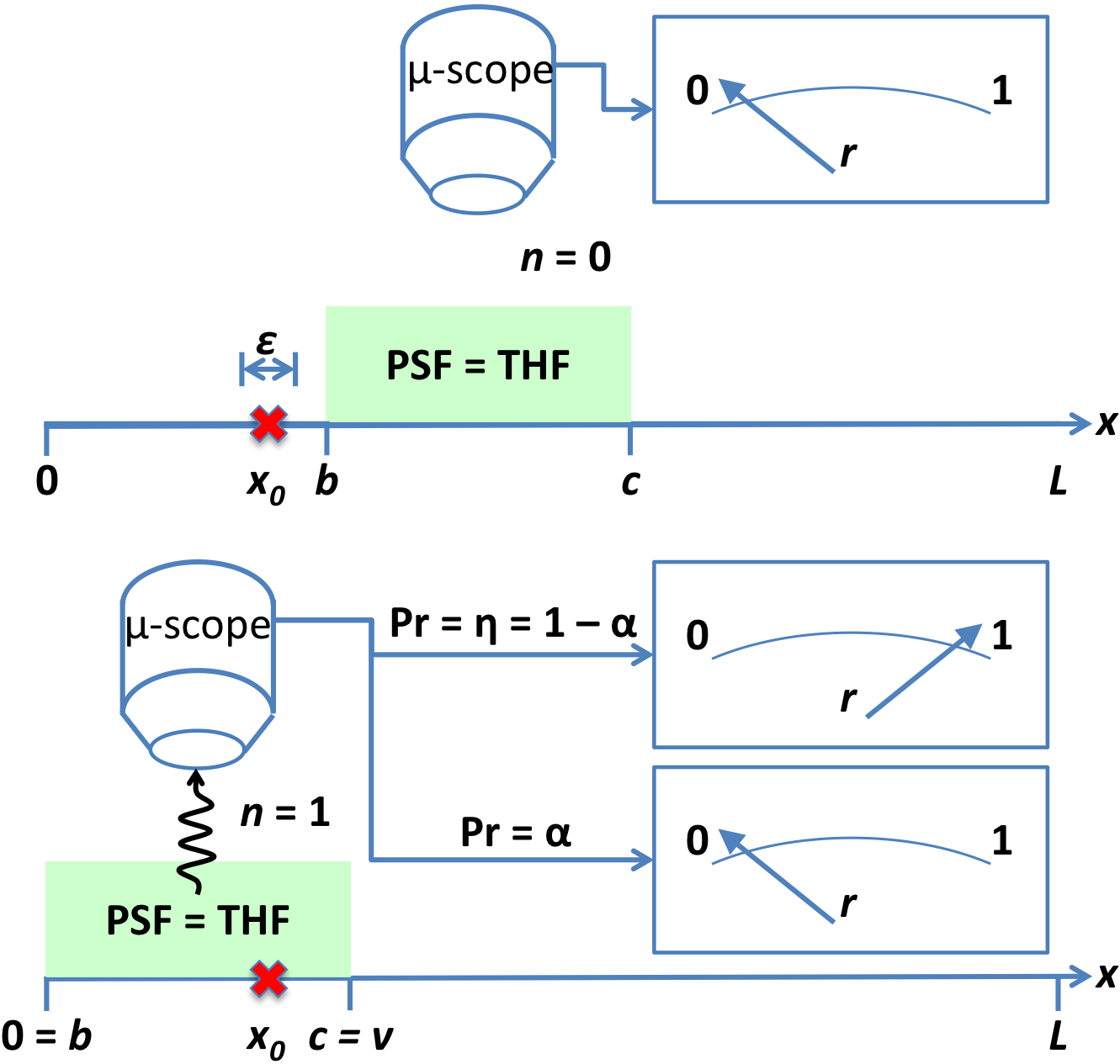}
\caption{Microscopy as a game: determining a fluorescing target's position (red cross) $x_{0}$ on $x\in\left[0,L\right]$ within $\epsilon$.  Top: microscope ($\mu$-scope) top-hat point-spread function (green-shaded area) not aligned with target ($x_{0}$~$<$~$b$~$<$~$c$); no photon generated ($n = 0$); measurement result guaranteed $r = 0$.  Bottom: target within microscope PSF ($b<x_{0}<c$); photon generated ($n = 1$).  Possibility of photon loss $\alpha$, leading to probabilistic measurement result ($r \in\left\{0,1\right\}$).  Here, $c=v$ as discussed in the main text; also $v$ and $\alpha$ are uniquely specified in each measurement.  The game ends when the distribution for $x$ is bounded within any distance $\epsilon$ not necessarily centred on the true location $x_{0}$, which is never found.}
\label{fig:schem}
\end{figure}

We wish to determine one (single-photon-emitting) fluorescent target's position $x_{0}$ in some continuous 1D domain $x\in[0,L]$ to within some precision $\epsilon$ (or alternatively, to better than some threshold resolution $\epsilon^{-1}$).  The precision may be considerably larger than the target's physical size, which is always on the atomic to few-nanometre scale.  The average number of queries required should be minimised, and the result achieved must be 100\% fidelitous.  We assume the microscope illuminates and collects from a region defined by a single top-hat function (THF) with arbitrarily controllable boundaries, as this is the simplest approximation we can make.  This is equivalent to a query in the R\'enyi-Ulam game over the THF.  Queries provoke $n$ fluorescence signal photons from the target where $n\in\left\{0,1\right\}$.  $n = 1$ \textit{iff} the target is in the non-zero region of the THF, otherwise $n = 0$.  These queries are formally equivalent to R\'enyi-Ulam interval queries, \textit{e.g.}, ``Is $b<x_{0}<c$?'', for some threshold values $b$ and $c$ (see Fig. \ref{fig:schem}).  (Note, however, that under some circumstances -- as described in Sec. \ref{sec:proposals} -- these are also equivalent to comparison queries.)

Microscopes often exhibit photon loss mechanisms, including finite numerical aperture, attenuation through optical fibres, and detector inefficiencies.  Such system-wide losses are usually (multiplicatively) characterised through an overall efficiency, $\eta$.  This parameter describes the chance of collecting a photon if the target is illuminated by the peak value of the point-spread function, and since photons are usually collected through the same PSF, if collection occurs through the peak value also.  Most microscopes have non-flat PSFs, often modelled with Gaussian functions.  Thus, the true efficiency is $\eta_{0}=\eta\times{\rm PSF}^{2}\left(x_{0}\right)$, if we normalise the peak value to 1.  For simplicity, we treat the PSF as a flat top-hat, and hence the true efficiency is simply $\eta$ (if the target is illuminated).  Being concerned primarily with photon loss, we define it as the complement of collection events, $\alpha_{0} = 1-\eta$, regardless of the origin of the loss.

Photon loss thus ocurrs with independent probability $\alpha_{0}$~$\in$~$\left[0,1\right]$.  This loss is the genesis mechanism of the half-lie.  Surviving photons are collected and counted; their number is the reported measurement result $r\in\left\{0,1\right\}$.  We formally connect target presence with a R\'enyi-Ulam ``yes'', its absence with a ``no'', and allow half-lies on ``yes'' states only.  \textit{I.e.}, a ``yes'' could be reported as a ``no'', but a ``no'' state can never be reported as a ``yes'' (see Table \ref{tab:connection}, or Eq. \ref{eq:physics}).  The game is immediately extensible in multiple commuting spatial dimensions; one plays a separable, independent game in each such dimension.

\begin{table}
\begin{ruledtabular}
\begin{tabular}{c|cc}
System & State	& Possible responses\\
\hline
\multirow{2}{*}{R\'enyi-Ulam} &No	& No\\
& Yes	& No or Yes\\
\hline
\multirow{2}{*}{Microscopy} & Absent	& $r = 0$\\
& Present	& $r\in\left\{0,1\right\}$\\
\end{tabular}
\caption{Possible responses in R\'enyi-Ulam game and microscopy experiment, illustrating the connection between the experiment and the game.  Because the ``No'' and $r = 0$ responses appear for both states, the possibility of a half-lie must be considered when one of these responses is obtained.  Conversely, a ``Yes'' or $r = 1$ response is guaranteed fidelitous.}
\label{tab:connection}
\end{ruledtabular}
\end{table}

Since we have probabilistic, random photon loss, we explicitly play a statistical, rather than a combinatorial, R\'enyi-Ulam game.  The number of loss events is theoretically unbounded.  Therefore, optimal solutions to combinatorial R\'enyi-Ulam games as described above do not exist here, since they describe scenarios where the number of lies (or half-lies) is known \textit{a priori}.    

For cases where up to a certain total fraction of the queries may be lied to (in any order, and in adversarial fashion), it has been shown that an $O\left(\log_{2} n\right)$ questioner's solution exists for lie-rates $<$ 1/3 (Ref. [\onlinecite{Spencer92}]), whilst for rates above the responder can always win.  The authors of [\onlinecite{Spencer92}] note that their result is also obtainable from the proofs of [\onlinecite{Rivest80}].  The complexity of the half-lie problem has been generally equated to that of the full-lie problem, for specified numbers of half-lies\cite{Rivest80}, and the number of necessary and sufficient queries for one half-lie has been solved.\cite{Cicalese00}  However, this still does not describe the probabilistic nature of the half-lies inherent to microscopy.  The combination of half-lies and the statistical variant does not appear to have been yet considered.  

Since confocal microscopes typically have probabilistic $\alpha_{0}>1/3$, but do not operate in adversarial fashion, we instead concern ourselves with minimising the \textit{average} number of queries.  Such an approach will aid experiments requiring many replications and/or variations, as are common in biology and medicine.


\section{Mathematical formalism}
\label{sec:maths}

We now formulate the game, with particular regard for the choice of measurement boundary (defined shortly) and the consequences of that choice for algorithm run time.  Random variables $X, N$, and $R$, are defined on the domains of $x$, $n$, $r$ (position, number of signal photons, measurement result); $A$ is the signal loss probability, on $\alpha\in\left[0,1\right]$, and $V$ is the turn-off boundary between the measured and non-measured regions, defined on $v\in\left[0,1\right]$.  Without loss of generality due to the invariance of Shannon information with a reordering of the distribution, we fix the turn-on boundary at the leftmost point with PDF $>0$ for convenience.  This also further specifies the query type as a comparison question.

The physics of our system allows us to put constraints on their behaviour:
\begin{equation}
\begin{split}
\mathbb{P}_{N|V,X}\left(n|v,x\right) &= \theta\left(v-x\right)\delta_{n,1} + \theta\left(x-v\right)\delta_{n,0}, {\rm ~and}\\
\mathbb{P}_{R|N,A}\left(r|n,\alpha\right) &= \delta_{n,0}\delta_{r,0} + \delta_{n,1}\left[\alpha\delta_{r,0}+\left(1-\alpha\right)\delta_{r,1}\right]
\end{split}
\label{eq:physics}
\end{equation}
where $\theta$ is the Heaviside step function, and $\delta_{i,j}$ is the Kronecker delta function as the domains $n$ and $r$, of $N$ and $R$ respectively, are discrete.  Since we only probe emitters left of $v$, we can only excite emissions left of $v$; the receipt of a photon necessarily requires its emission, and is sometimes blocked by signal loss.  For conciseness, we henceforth omit the probabilities' domains.

As the domain $v$ is continuous, we set $\mathbb{P}_{V}~=~\delta\left(v-v_{0}\right)$, using the Dirac delta function, and assuming a perfectly known boundary, $v_{0}$.  Also, $\mathbb{P}_{A}~=~\delta\left(\alpha-\alpha_{0}\right)$, a perfectly known false negative rate, $\alpha_{0}$.  Further, although the target has a unique position $x_{0}$, this is unknown; therefore we have a uniform prior over $X$, such that $\mathbb{P}_{X} = 1/L$.

Since $A$ and $V$ are uniquely determined,
\begin{equation}
\begin{split}
\mathbb{P}_{A|N,V,X} &= \mathbb{P}_{A};\\
\mathbb{P}_{V|N,X} &= \mathbb{P}_{V}.
\label{eq:avgnx}
\end{split}
\end{equation}

\begin{equation}
\begin{split}
{\rm Also,~}\mathbb{P}_{N,V,X} &= \mathbb{P}_{N|V,X}\mathbb{P}_{V|X}\mathbb{P}_{X},\\
&= \mathbb{P}_{N|V,X}\mathbb{P}_{V}\mathbb{P}_{X}, {\rm and}\\
\mathbb{P}_{N,X} &= \int_{V}\mathbb{P}_{N|V,X}\mathbb{P}_{V}\mathbb{P}_{X}dv.
\end{split}
\label{eq:nx}
\end{equation}

We now consider that when we attempt to detect photons, we are making a quantum measurement of the field.  Since the position of the emitter and measurement boundary are determined before the Markov process of measuring (collapsing) the state and reporting the outcome, we can deconditionalise thus:
\begin{equation}
\mathbb{P}_{R|N,A,V,X} = \mathbb{P}_{R|N,A}.
\label{eq:quantum}
\end{equation}

We form the full joint by:
\begin{equation}
\begin{split}
\mathbb{P}_{N,A,V,X,R} &= \mathbb{P}_{R|N,A,V,X}\mathbb{P}_{N,A,V,X}\\
&= \mathbb{P}_{R|A,V,N,X}\mathbb{P}_{A,V|N,X}\mathbb{P}_{N,X}\\
&= \mathbb{P}_{R|A,V,N,X}\mathbb{P}_{A|V,N,X}\mathbb{P}_{V|N,X}\mathbb{P}_{N,X}
\end{split}
\label{eq:joint}
\end{equation}

Substituting Eqs. \ref{eq:physics}--\ref{eq:quantum} in Eq. \ref{eq:joint} and marginalising gives
\begin{equation}
\begin{split}
\mathbb{P}_{X,R} &=\sum_{N}\int_{A,V}\mathbb{P}_{N,A,V,X,R}d\alpha dv\\
&=\frac{1}{L}\left[\delta_{r,0}\theta\left(x-v_{0}\right)+\alpha_{0}\delta_{r,0}\theta\left(v_{0}-x\right)\right.\\
&\left.\qquad+\left(1-\alpha_{0}\right)\delta_{r,1}\theta\left(v_{0}-x\right)\right],{\rm ~and}\\
\mathbb{P}_{R}&=\int_{X}\mathbb{P}_{X,R}dx\\
&=\frac{1}{L}\left[\delta_{r,0}\left(L-v_{0} + \alpha_{0}v_{0}\right) + \delta_{r,1}\left(1-\alpha_{0}\right)v_{0}\right]
\end{split}
\end{equation}

\begin{figure*}[tb!]
\centering
\includegraphics[angle=0, width=0.99\linewidth]{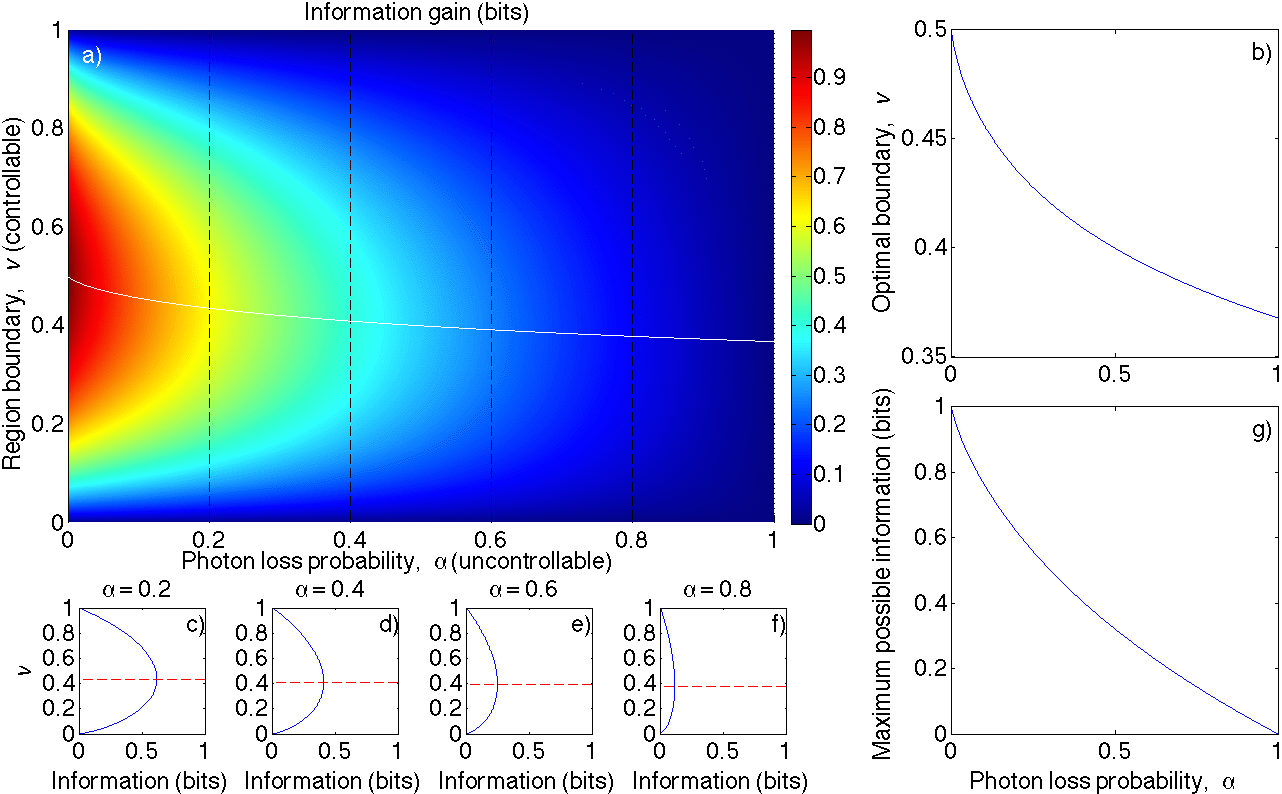}
\caption{Optimal measurement domain strategy per independent measurement on $[0,L]$ in the presence of photon loss half-lies: a) false-colour plot of information gained as a function of false negative rate $\alpha$ and region boundary $w$, with particular values of $\alpha$ marked (black dashed lines), and optimal region boundary placement shown as a function of $\alpha$ (white line); b) focus on optimal boundary placement (white line from a); c)-f) line slices of a) showing the information gain for each boundary placement at $\alpha\in\left\{0.2,0.4,0.6,0.8\right\}$ (black lines in a); and g) maximal information possible per measurement shown as a function of $\alpha$.}
\label{fig:alphav}
\end{figure*}

Now, using the definitions of Shannon information:
\begin{equation}
\begin{split}
H\left(X\right) &= -\int_{X}\mathbb{P}_{X}\log_{e}\left(\mathbb{P}_{X}\right)~dx,\\
{\rm and~}H\left(X|R\right) &=~\sum_{R}\int_{X}\mathbb{P}_{X,R}\log_{e}\left(\frac{\mathbb{P}_{R}}{\mathbb{P}_{X,R}}\right)~dx,
\end{split}
\end{equation}
we can calculate the mutual information:
\begin{equation}
\begin{split}
I\left(X:R\right) &= H\left(X\right) - H\left(X|R\right)\\
&= \log_{e}\left(\frac{1}{L}\right)-\frac{1}{L}\left[\left(1-\alpha_{0}\right)v_{0}\log_{e}\left(v_{0}\right)\right. \\
&\left.\qquad+ \left(L-v_{0}+\alpha_{0}v_{0}\right)\log_{e}\left(L-v_{0}+\alpha_{0}v_{0}\right)\right. \\
&\left.\qquad- \alpha_{0}v_{0}\log_{e}\left(\alpha_{0}\right)\right]
\end{split}
\label{eq:mutinf}
\end{equation}
which we maximise to find the condition:
\begin{equation}
v_{0} = \frac{L\alpha_{0}^{\frac{\alpha_{0}}{1-\alpha_{0}}}}{1+\left(1-\alpha_{0}\right)\alpha_{0}^{\frac{\alpha_{0}}{1-\alpha_{0}}}},
\label{eq:limit}
\end{equation}
shown in Fig. \ref{fig:alphav}.  As expected, since this limit is formally equivalent to the canonical binary search problem, $\lim_{\alpha_{0}\rightarrow 0}v_{0} = 1/2$ and is undefined for $\alpha_{0}=1$; however, the limiting value as $\alpha_{0}\rightarrow1$ is $v_{0}\rightarrow1/e$.  The optimal measurement domain in the presence of photon loss (half-lies) is therefore less than $L/2$.

\section{Search strategies}
\label{sec:search}


\subsection{Proposals}
\label{sec:proposals}

A greedy strategy for target searches would be to maximise the information gain for each subsequent measurement.  However, any search over a region that is not a simple dividend of the overall region will be difficult to update for a null result, $r = 0$.  (It would also return the query type to interval questions.)  We therefore consider simple heuristic approaches, whereby the domain is split into $q$ subdomains which are explored sequentially until the target is explicitly located, whence the split (into $q$ subdomains) recurs.  Note that the final split in each heuristic is still into $q$ subdomains, even if fewer would suffice to meet the criterion of precision $\epsilon$, maintaining the heuristics' simplicity.  We develop and contrast four such approaches: bi-, tri-, and tetra-sectioning with verification, and the limiting extension of these to sequentially scanning each of the $\epsilon^{-1}$ subdomains.

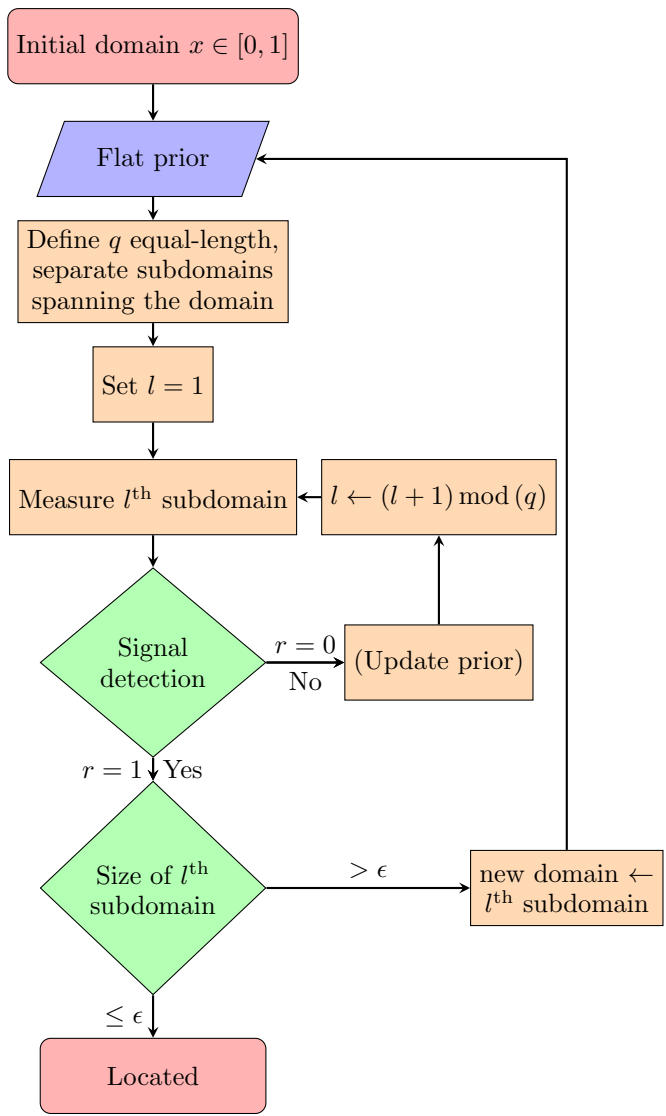
\begin{figure}[tb!]
\begin{tikzpicture}[node distance = 2cm]
\node (start) [startstop] {Initial domain $x\in\left[0,1\right]$};
\node (io1) [io, below of = start, yshift=0.5cm] {Flat prior};
\node (def1) [process, below of = io1, yshift=0.5cm]{Define $q$ equal-length,\\separate subdomains\\spanning the domain};
\node (proc1) [process, below of = def1, yshift=0.5cm] {Set $l = 1$};
\node (meas1) [process, below of = proc1, yshift=0.5cm] {Measure $l^{{\rm th}}$ subdomain};
\node (dec1) [decision, below of = meas1, yshift=-0.2cm] {Signal\\detection};
\node (proc2) [process, right of = dec1, xshift=1.8cm] {(Update prior)};
\node (proc3) [process, above of = proc2, yshift=0.2cm] {$l\leftarrow\left(l + 1\right){\rm mod}\left(q\right)$};
\node (dec2) [decision, below of = dec1, yshift = -1.0cm] {Size of $l^{{\rm th}}$ \\subdomain};
\node (proc4) [process, right of = dec2, xshift = 3.5cm] {new domain $\leftarrow$ \\ $l^{{\rm th}}$ subdomain};
\node (output) [startstop, below of = dec2, yshift = -0.5cm] {Located};
\draw [arrow] (start) -- (io1);
\draw [arrow] (io1) -- (def1);
\draw [arrow] (def1) -- (proc1);
\draw [arrow] (proc1) -- (meas1);
\draw [arrow] (meas1) -- (dec1);
\draw [arrow] (dec2) -- node[anchor=east] {$\leq \epsilon$} (output);
\draw [arrow] (dec2) -- node[anchor=south] {$> \epsilon$} (proc4);
\draw [arrow] (dec1) -- node[anchor=east] {$r =1$} (dec2);
\draw [arrow] (dec1) -- node[anchor=west] {Yes} (dec2);
\draw [arrow] (dec1) -- node[anchor=south] {$r = 0$} (proc2);
\draw [arrow] (dec1) -- node[anchor=north] {No} (proc2);
\draw [arrow] (proc2) -- (proc3);
\draw [arrow] (proc3) -- (meas1);
\draw [arrow] (proc4) |- (io1);
\end{tikzpicture}
\caption{$q$-sectioning algorithm requiring detection to proceed.  Update prior step not formally required.  We choose (without loss of generality) to index subdomains from the left boundary.  $l$ is an index for the $q$ subdomains required at any step of $q$-sectioning.  The algorithm completes when the final subdomain is smaller than the precision required, $\epsilon$.}
\label{fig:flow}
\end{figure}

The obvious choice for an heuristic is bisectioning, the limiting case for zero half-lie rate.  (Basic bisectioning is equivalent to binary search in this limit.)  Here, we define an approach where null measurements simply unbalance the PDF, and following measurements are undertaken on the other half; \textit{i.e.}, na\"ively rastering between only two equal-sized subdomains.  This cycle repeats until positive verification of target presence occurs by a measurement $r=1$; the problem is then recursively reposed within the successful half (Fig. \ref{fig:flow}, for $q=2$).  We call this process ``bisectioning with verification''.  Note that by insisting on positive verification, we cannot take advantage of all the available information in the low-half-lie limit.  On average, this process dictates 1.5 measurements per level of enquiry, due to the equal probabilities of the target being or not being in the first-examined subdomain; \textit{i.e.}, for precision $\epsilon$, $1.5\lceil\log_{2}\left(\epsilon^{-1} \right)\rceil$ queries are required.

Here, we extend the definition of bisectioning with verification to define a new family of heuristics: $q$-sectioning with verification.  $q$ equal subdomains are queried in turn until positive verification is obtained.  The average number of measurements will be $\bar{m}_{q,\epsilon,\alpha_{0}}$; for half-lie-free search this is: 
\begin{equation}
\bar{m}_{q,\epsilon,0}=\frac{\left(q+1\right)}{2}\lceil\log_{q}\left(\epsilon^{-1}\right)\rceil.
\label{eq:qsectioning}
\end{equation}

Now we consider trisectioning with verification; we have $\bar{m}_{3,\epsilon,0} = 2\lceil\log_{3}\left(\epsilon^{-1}\right)\rceil$.  Due to our high probability of signal loss, we expect $\bar{m}_{3,\epsilon^{-1},\alpha_{0}\neq0}$ to be considerably higher than $\bar{m}_{3,\epsilon^{-1},0}$.  Note that this is explicitly not the ternary search of R\'enyi-Ulam games, defined as a special case of $q$-ary search.\cite{Soskov89,Muthukrishnan94}  That refers instead to a similar, but subtly different process in which the Responder indicates which of the $q$ subdomains holds the target.  A good example problem of that type is finding a heavier coin by balance weighing.\cite{Cicalese13}  Neither is it the ternary search of computer science or learning, which maximises unimodal functions by evaluating their value at two intermediate points,\cite{Dobkin91,Salehin10} or classifies and sorts data.\cite{Bentley97}

We also extend this idea to tetrasectioning, which is analogous to bi- and tri-sectioning but partitions the surviving search space into four equal subdomains at any level of measurement.  Similarly, the average number of measurements for half-lie-free search will be\qquad\qquad $\bar{m}_{4,\epsilon^{-1},0}=2.5\lceil\log_{4}\left(\epsilon^{-1}\right)\rceil$.

The penultimate approach we consider is using precisely as many equal sections as we require for the desired precision, \textit{i.e.}, $\epsilon^{-1}$ subdomains, and rastering over these subdomains until positive verification is obtained.  We call this approach ``rastering with verification''.  This uses $\bar{m}_{q=\epsilon^{-1},\epsilon^{-1},0}=0.5\left(\epsilon^{-1} + 1\right)\log_{\epsilon^{-1}}\left(\epsilon^{-1}\right)$ measurements.  For any appreciable value of $\epsilon^{-1}$, this is prohibitively large compared to the previous strategies.  We therefore summarily dismiss rastering with verification as unfeasible (except for small $\epsilon$, where it approaches the other strategies) and do not consider it further in this work.

Finally, we consider rastering as is commonly performed.  Here, the domain is continuously scanned from left to right, with the THF having width $\epsilon$.  This is used instead of dwelling on each subdomain in turn to avoid ring-down of the equipment after stopping and any consequent dark time to allow its mitigation.  The scan rate is slow for several reasons, including shot noise suppression, and convenience in automation by standardising the process for every pixel.  We can, however, model this approach as physically dwelling on each subdomain in turn for a set number of queries before moving on.  For a dwell time corresponding to $\gamma$ queries, this approach will require $\gamma/\epsilon$ measurements; $0.5\gamma/\epsilon$ on average if the scan is adaptively implemented and will cease after locating the target.  Not only does this approach again require a far larger number of measurements for any appreciable $\epsilon^{-1}$, but the inefficiency is compounded by the dwell parameter, $\gamma$, which must be at least 1.  Once more, we dismiss this strategy and ignore it henceforth.

We note in each of these heuristics that since the initial domain PDF is uniform, the entropy of an initial query on an edge subdomain is the same as one on an internal subdomain (of equivalent length).  Therefore, the query types, though often of interval form, are informationally equivalent to comparison queries.  Similar arguments can be made for the following measurements until a new mapping within a subdomain occurs and the reasoning recurs.


\subsection{First-pass analysis}
\label{sec:analysis}

To estimate the optimal average number of queries needed to locate a target within $\epsilon$ for specific $\alpha_{0}$, we consider the mutual information (Eq. \!\!\!\ref{eq:mutinf}) delivered by a hypothetical optimal first measurement (Eq. \!\!\ref{eq:limit}).  We approximate all consequent measurements with this value.  Achieving the desired precision corresponds to acquiring $\log_{2}\left(\epsilon^{-1}\right)$ bits of information, and requires an estimated $\lceil\left[\log_{2}\left(\epsilon^{-1}\right){\rm ~bits}\right]/\left[I\left(X:R\right)|_{\alpha_{0},v_{0}}{\rm ~bits/query}\right]\rceil$ queries.

We simulated one million random emitter positions, using $\alpha_0=0.99$, and subjected each to bi-, tri-, and tetra-sectioning with verification as described above, to an arbitrarily selected precision of $\epsilon = 10^{-10}$.  (This high precision displays the algorithmic speedup well; microscopy generally deals with $\epsilon>10^{-6}$, or the approximate ratio of a confocal spot diameter to the length of a 96-well plate.)  The measurement regions were curtailed only after positive detection, otherwise cycling through the 2, 3, or 4 sections available.  For $\alpha_{0}=0.99$, this dictates an estimated $\approx6240$ ``optimal'' queries.

\begin{figure}[tb!]
\centering
\includegraphics[angle=0, width=0.99\linewidth]{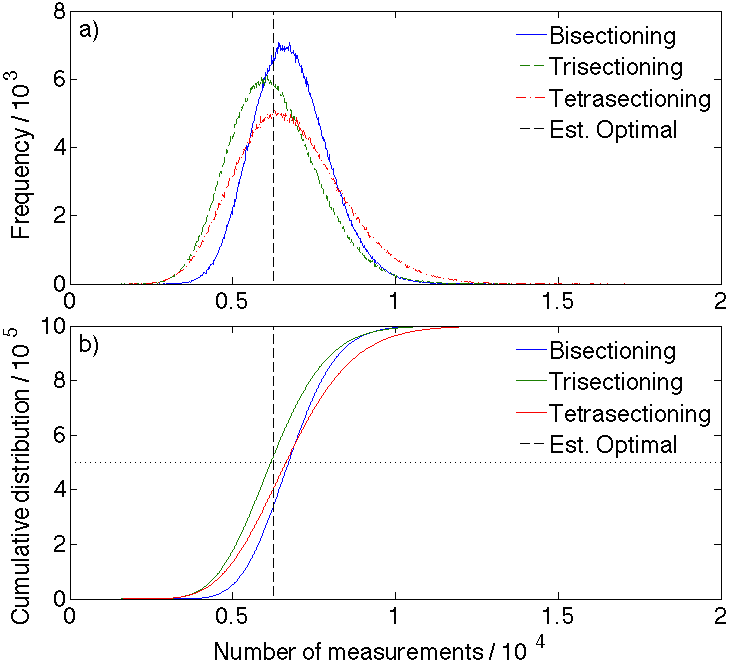}
\caption{Results of 1,000,000 Monte Carlo trials with $\alpha_{0}$~$=$~$0.99$ for each sectioning strategy: a) Histograms of search heuristic results indicating the mode of each strategy, and estimated optimal average result, and b) cumulative distributions of same, indicating median values.  Note: the trisectioning trace is to the left of both bi- and tetra-sectioning for almost the entire number of trials.  Also note the estimated optimal average (dotted black line), similar but not equal to the trisectioning median (intersections with dashed black line).}
\label{fig:comparison}
\end{figure}

Figure \ref{fig:comparison}a shows a histogram of the Monte Carlo results: the mode [Mo()] of each stratagem is apparent, with Mo(trisectioning) $<$ Mo(tetrasectioning) $<$ Mo(bisectioning).  The trisectioning mode is also smaller than the estimated optimal average number of measurements; the semi-infinite nature of the data space allows for an average value pulled right of the mode.

Fig. \ref{fig:comparison}b is the cumulative density function for each approach, from which the median is extractable where each curve crosses the horizontal dotted line.  Again, the same ordering is evident, with Median(trisectioning) falling just below the estimated optimal average number.

The distribution means are displayed in the first row of Table \ref{tab:meansearch}, and are consistent with the other measures of central tendency, except that the trisectioning mean is now slightly higher than the estimated optimal average.  In addition, the mean number of queries required was evaluated for several different half-lie rates, and form the remainder of Table \ref{tab:meansearch}.  Similar estimates for the optimal behaviour based on $I\left(X:R\right)|_{\alpha_{0},w}$ are also provided.

\begin{table}
\begin{ruledtabular}
\begin{tabular}{c|cccc}
$\alpha_{0}$	& $\overline{{\rm Bi-}}$	& $\overline{{\rm Tri-}}$	& $\overline{{\rm Tetra-}}$	& Est.\\
			& sectioning			& sectioning			& sectioning	& optimal\\
\hline
0.99	& 6783.13	& 6275.92	& 6771.67	& 6240	\\
0.5	& 119.00	& 105.02	& 110.49	& 104	\\
0.3	& 80.15	& 69.00	& 71.64	& 66	\\
0.1	& 58.55	& 49.01	& 50.06	& 44	\\
1/18	& 54.99	& 45.71	& 46.49	& 40	\\
1/35	& 53.00	& 43.85	& 44.60	& 37	\\
1/36	& 52.94	& 43.80	& 44.44	& 37	\\
0.01	& 51.69	& 42.64	& 43.18	& 35	\\
0.001	& 51.07	& 42.06	& 42.57	& 34	\\
0	& 51.00& 42.00	& 42.50	& 34
\end{tabular}
\caption{Average number of measurements (for $10^{6}$ trials) to locate an emitter to one of  $10^{10}$ subdomains  using various heuristic search strategies and an estimate of the optimal average value obtained from Eqs. \ref{eq:mutinf} and \ref{eq:limit} (shown in Fig. \ref{fig:comparison}g).}
\label{tab:meansearch}
\end{ruledtabular}
\end{table}

The requirement of absolute knowledge (zero PDF outside the final interval), by positive verification of presence through signal capture, inflicts a penalty on the sectioning strategies.  As mentioned above, 33.2 bits of information are required: at zero half-lie rate, this dictates 34 measurements in a perfect scheme.  The final row of Table \ref{tab:meansearch} shows that none of the sectioning strategies performs particularly well in that case, with bisectioning using on average 50\% extra queries.  In contrast, tri- and tetra-sectioning only require 24\% extra queries.  Neither uses the (positive) information that can be derived from null results, which faithfully indicate the target's absence.  The ignorance and disuse of the full information available by the insistence on positive verification translates as an extra cost in the number of measurements required.

It is instructive to consider the lower (non-trivial) limit of half-lie rates, where the most likely outcome is one half-lie.  Assuming the 34 necessary plus one erroneous measurements, one half-lie is most likely for $1/36\leq\alpha_{0}\leq 1/18$, with the expected value of one half-lie for $\alpha_{0}=1/35$.  Rivest \textit{et~al.}\cite{Rivest80} proved that finding a target in $k$ discrete subdomains , when up to $E$ responses to comparison questions may be erroneous \textit{iff} the truth is of one specified type (\textit{e.g.}, less than), requires $Q\geq \lceil\log_{2}\left(k\right) + E\log_{2}\log_{2}\left(k\right) + O\left[E\log_{2}\left(E\right)\right]\rceil$ comparison questions.  For our example case, where $E=1$ and $k=10^{10}$, $E\log_{2}\left(E\right)=0$, requiring $Q\geq\lceil\sim38.2\rceil=39$ queries.  From Table \ref{tab:meansearch}, the trisectioning heuristic performs on average within 11\% of the Rivest \textit{et al.} bound within these limits, and outperforms the bi- and tetra-sectioning approaches.

Not only does bisectioning fail to be optimal (even for zero half-lies), trisectioning outperforms it at every half-lie rate as well.  From Table \ref{tab:meansearch} we can determine that the penalty for using bi- instead of tri-sectioning is 8--21\%, depending on the photon loss rate -- and worst for low loss rate (where the insistence on positive verification is most deleterious).  There is therefore an immediate, quantifiable benefit to preferring the trisectioning strategy over bisectioning.  Tetrasectioning has a similar, albeit smaller, penalty, and is nearly as efficient as trisectioning for low half-lie rates.

Although for low photon loss, the trisectioning heuristic requires more measurements than the estimated optimal information use, its performance improves markedly as losses increase.  For typical confocal microscope conditions ($\alpha_{0}\approx0.99$), trisectioning requires on average only 1\% more measurements, and beats bisectioning by 10\%.

Returning to the original context of confocal/multiphoton microscopy, one major point of difference is that here, we explicitly recognise that once a subdomain is searched and a null result obtained, the target is more likely elsewhere in the full domain.  In contrast, conventional approaches dwell on each pixel, effectively performing a number of measurements (as defined by our approach) several orders of magnitude larger than the one performed here before moving the PSF.  Such processes are information inefficient.

The second, and more impactful, difference is that conventional microscopy \textit{always} searches the same amount of area per query, whereas the sectioning approaches resize the PSF after verification.  This is the origin of the change from linear to logarithmic behaviour -- if the local intensity of the probe beam within the PSF can be maintained for diffuse measurements.


\subsection{Exploration for other precisions}
\label{sec:otherres}


\subsubsection{Zero half-lie rate}
\label{sec:nolies}

The selection above of $\epsilon^{-1}=10^{10}$ was somewhat arbitrary, and it is natural to question whether the performance of the trisectioning algorithm was due to the nearness of $10^{10}$ to a power of three (and distance from powers of two and four).

\begin{figure*}[tb!]
\centering
\includegraphics[angle=0, width=0.99\linewidth]{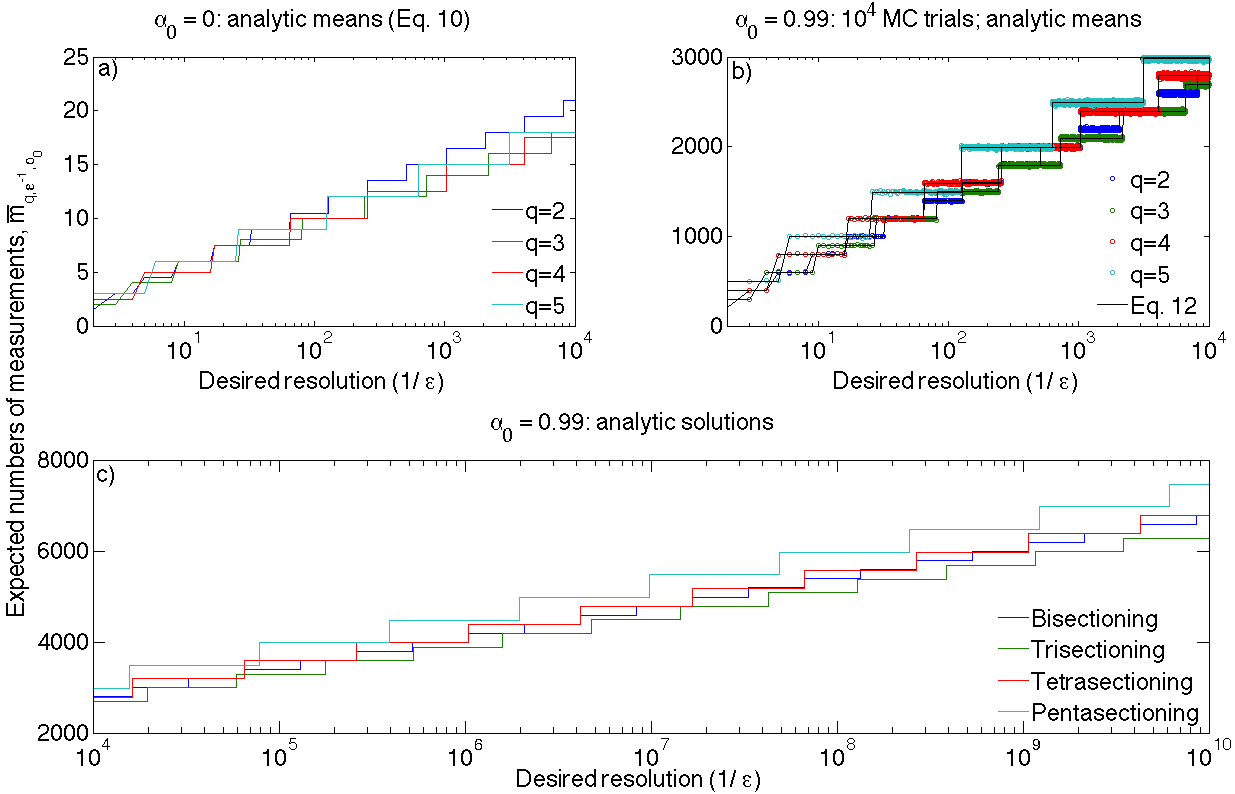}
\caption{Average numbers of searches required to find a target for various $\epsilon$, given particular heuristics and $\alpha_{0}$.  a) Directly calculated exact means for $\alpha_{0}=0$; b) Mean results of running 10$^{4}$ Monte Carlo trials for each value of $\epsilon^{-1}\in\left\{2,3, ...,10^{4}\right\}$ and $\alpha_{0}=0.99$; c) direct calculation of expected means for the same conditions as b); d) extension of direct calculation of c) for all $\epsilon^{-1}\in\left\{10^{4}+1,10^{4}+2, ...,10^{8}\right\}$.}
\label{fig:naivestupid}
\end{figure*}

To explore this concept, we propose the following thought experiment: let us set the half-lie rate to $\alpha_{0}=0$.  Now, there is no probabilistic component to the average number of searches required to find a target in $q$ subdomains; the average number is simply $\frac{q}{2}$, based on the random placement of the target within the initial domain.  We can therefore easily compute the expected number of searches under a given heuristic as $\frac{q}{2}\lceil\log_{q}\left(\epsilon^{-1}\right)\rceil$.  Performing this for $q\in\left\{2,3,4,5\right\}$ and $\epsilon^{-1}\in\left\{2,3,4,...,10^{6}\right\}$, we compute Fig. \ref{fig:naivestupid}a.

It is quickly apparent that trisectioning and tetrasectioning dominate a search for minimum-search strategies.  Further investigation shows that tetrasectioning is actually the most frequent winner, for 611612 of the considered values of $\epsilon^{-1}$.  It is unique for 569970 of these cases.   See Table \ref{tab:naive} for more detail.

\begin{table}
\begin{ruledtabular}
\begin{tabular}{c|cc}
$q$ & \multicolumn{2}{c}{Solution type} \\
& Optimal & Unique\\
\hline
2 & 2 & 1\\
3 & 415,229 & 246,610\\
4 & 611,612 & 569,970\\
5 & 142,433 & 12,634\\
6 & 1,500 & 11\\
7 & 11 & 1\\
8 & 1 & 0\\
10 & 0 & 0\\
100 & 0 & 0\\
1,000 & 0 & 0\\
\end{tabular}
\caption{Performance of various $q$-sectioning search heuristics for 1/$\epsilon \in \left\{2, 3, ..., 10^{6}\right\}$ and $\alpha_{0}=0$.  Performance is described as optimal where no $q$-sectioning heuristic solution exists for that $\epsilon^{-1}$ with fewer measurements, and unique if no similarly optimal solution for that $\epsilon^{-1}$ exists with different $q$.}
\label{tab:naive}
\end{ruledtabular}
\end{table}

Strategies with $q > 8$ are never efficient, hence our comments about rastering as generally performed.  Strategies with $5\leq q\leq 8$ are efficient only in limited special cases.  Choosing $q \in \left\{3, 4\right\}$ offers optimal behaviour for 987,353 of the cases considered.


\subsubsection{Realistic microscopy parameters}
\label{sec:somelies}

The situation becomes more interesting for non-zero $\alpha_{0}$, and yet harder to obtain extensive data for.  Since non-zero $\alpha_{0}$ implies some probabilistic behaviour, we again performed many Monte Carlo simulations to obtain the equivalent of one datum from Fig. \ref{fig:naivestupid}a, but for $\alpha_{0}=0.99$.  We chose $\epsilon^{-1}=10^{6}$ as equivalent to a search across a microscope slide or well plate (of order 10 cm) down to the diffraction limit of light (of order 100 nm).  It is also conveniently substantially closer to powers of 2 and 4 than 3, as given by our metric function for the effective nearness of $\epsilon^{-1}$ to a power of $q$: 
\begin{equation}
F\left(q,\epsilon\right) = 0.5 - \left\|0.5 Ð \left[\log_{q}\left(\epsilon^{-1}\right) - \lfloor\log_{q}\left(\epsilon^{-1}\right)\rfloor\right]\right\|
\end{equation}
A value of $F=0$ means that $\epsilon^{-1}$ is an exact match to a power of $q$; $F=0.5$ is the greatest distance attainable.

We find that trisectioning is still competitive (Table \ref{tab:million}).  We have also included pentasectioning for completeness; it performs poorly compared to the others, despite $10^{6}$ being a similar distance from a power of five as from a power of three.  Bi-and tetra-sectioning, which have similar $F$ values, have drastically different modes.  Clearly, the relative nearness to a power of $q$ is not the only contributing factor to the performance of the heuristics.

\begin{table}
\begin{ruledtabular}
\begin{tabular}{c|cccc}
$q$ & $F\left(q,\epsilon\right)$ & Mean & Median & Mode \\
\hline
2 & 0.068 & 3990.2 & 3924 & 4157\\
3 & 0.425 & 3887.4* & 3789* & 3674\\
4 & 0.034* & 3983.5 & 3853 & 3458*\\
5 & 0.416 & 4482.8 & 4318 & 3657\\
\end{tabular}
\caption{Measures of central tendency from results of Monte Carlo trials as per Fig. \ref{fig:comparison}; 10$^{6}$ trials for $\alpha_{0}=0.99$.  The lowest column values are denoted by asterisks (*).}
\label{tab:million}
\end{ruledtabular}
\end{table}

We explored a limited subset of $\epsilon^{-1}$ to the same rigour (Table \ref{tab:more}).  Here, pentasectioning occasionally performs comparably to the other heuristics, but mostly considerably poorer than the best of them.  It is notable that the heuristic with minimal distance is most often not the one with the lowest mean number of measurements.

\begin{table}
\begin{ruledtabular}
\begin{tabular}{cc|cccc}
$\epsilon^{-1}$ & $q$ & $F\left(q,\epsilon\right)$ & Mean & Median & Mode \\
\hline
\multirow{4}{*}{$10^{3}$} &2 & 0.034 & 1994.7 & 1929 & 1813\\
& 3 & 0.288 & 2092.7 & 1994 & 1762\\
& 4 & 0.017* & 1992.0* & 1861* & 1568*\\
& 5 & 0.292 & 2489.5 & 2325 & 2016\\
\rule{0pt}{2.6ex}
\multirow{4}{*}{$3\times10^{3}$} &2 & 0.449 & 2394.6 & 2328 & 2108\\
& 3 & 0.288 & 2391.7 & 2292 & 2158\\
& 4 & 0.225 & 2389.7* & 2257* & 1890\\
& 5 & 0.025* & 2429.5 & 2329 & 1834*\\
\rule{0pt}{2.6ex}
\multirow{4}{*}{$10^{4}$} &2 & 0.288 & 2797.7 & 2716 & 2641\\
& 3 & 0.384 & 2690.3* & 2591* & 2607\\
& 4 & 0.356 & 2787.8 & 2656 & 2359*\\
& 5 & 0.277* & 2989.1 & 2823 & 2628\\
\rule{0pt}{2.6ex}
\multirow{4}{*}{$3\times10^{4}$} &2 & 0.127* & 2992.8 & 2927* & 2821\\
& 3 & 0.384 & 2988.7* & 2980 & 2602\\
& 4 & 0.436 & 3188.9 & 3059 & 2565*\\
& 5 & 0.405 & 3845.9 & 3321 & 3330\\
\rule{0pt}{2.6ex}
\multirow{4}{*}{$10^{5}$} &2 & 0.390 & 3391.9 & 3326 & 3145\\
& 3 & 0.480 & 3288.8* & 3190* & 2947*\\
& 4 & 0.305 & 3584.9 & 3452 & 3417\\
& 5 & 0.153* & 3983.7 & 3820 & 3704\\
\rule{0pt}{2.6ex}
\multirow{4}{*}{$3\times10^{5}$} &2 & 0.195 & 3790.7 & 3725 & 3657\\
& 3 & 0.480 & 3587.0* & 3488* & 3071\\
& 4 & 0.097* & 3985.2 & 3852 & 3637\\
& 5 & 0.164 & 3984.9 & 3823 & 3160\\
\rule{0pt}{2.6ex}
\multirow{4}{*}{$10^{6}$} &2 & 0.068 & 3990.2 & 3924 & 4157\\
& 3 & 0.425 & 3887.4* & 3789* & 3674\\
& 4 & 0.034* & 3983.5 & 3853 & 3458*\\
& 5 & 0.416 & 4482.8 & 4318 & 3657\\
\end{tabular}
\caption{Measures of central tendency from results of Monte Carlo trials as per Fig. \ref{fig:comparison}; 10$^{6}$ trials for $\alpha_{0}=0.99$.  The lowest column/major row values are marked by asterisks.}
\label{tab:more}
\end{ruledtabular}
\end{table}


\subsubsection{General solution}
\label{sec:general}

It being impractical to perform 10$^{6}$ simulations for each value of $\epsilon^{-1}$, we match Fig. \ref{fig:naivestupid}a by lowering our standards.  We performed 10$^{4}$ Monte Carlo simulations for every value of $\epsilon^{-1}$ between 2 and 10$^{4}$, with $\alpha_{0}$ set to 0.99; the results can be see in Fig. \ref{fig:naivestupid}b.  We note that the individual heuristics still perform in step-wise semilogarithmic fashion (albeit with larger jumps); however, the pattern is significantly distorted from the zero-half-lie case.  Our observations regarding pentasectioning are borne out; in the main it performs the worst of these heuristics.  We also note the enhanced perfomance of the trisectioning heuristic over the tetrasectioning, as evidenced by its greater proportion of being the minimum solution -- especially at the right of the figure, where the logarithmic scale greatly compresses the data over $\epsilon^{-1}$.

The regularity of each heuristic's behaviour led us to propose a more general form for the mean number of measurements, based upon the frequentist interpretation of probabilities as the mean chance of outcome given exhaustively many trials:
\begin{equation}
\bar{m}_{q,\epsilon^{-1},\alpha_{0}} = \left[q\left(\frac{1}{1-\alpha_{0}}-1\right)+ \frac{q+1}{2}\right]\lceil\log_{q}\left(\epsilon^{-1}\right)\rceil .
\label{eq:behaviour}
\end{equation}
The first term in the prefactor recognises that in the long run, 1/(1-$\alpha_{0}$) trials must occur in the correct subdomain before a photon will be detected.  However, for each trial (except the final one), we incur the penalty of searching the other $q-1$ subdomains, necessitating $q\left(\frac{1}{1-\alpha_{0}}-1\right)$ searches per level before we expect to receive a signal on the next pass through.  On the final pass (per level), we have ($q$+1)/2 searches as before.  This total, of course, is then scaled by the logarithmic ceiling function, which dictates how many levels of search need be accomplished to sufficiently locate the target.  Of course, $\lim_{\alpha_{0}\rightarrow 0} \bar{m}_{q,\epsilon^{-1},\alpha_{0}} = \bar{m}_{q,\epsilon^{-1},0}$ from Eq. \ref{eq:qsectioning}.

Figure \ref{fig:naivestupid}b also presents a direct calculation (by Eq. \ref{eq:behaviour}) of the means previously estimated through Monte Carlo sampling.  By eye, it describes the behaviour well.  We extend its domain over more values of $\epsilon^{-1}$ than it is practical to simulate, even for a single $\alpha_{0}$, in Fig. \ref{fig:naivestupid}c.  Given the semi-logarithmic nature of the plot, trisectioning search is the optimal $q$-sectioning heuristic for the vast majority of $\epsilon^{-1}\in\left\{2,3,...,10^{10}\right\}$ (and $\alpha_{0}=0.99$).  Closer exploration reveals it to be the optimal $q$-sectioning heuristic for all $\epsilon^{-1}>2^{24}\sim1.7\times 10^{7}$ (and $\alpha_{0}=0.99$).  It is also optimal for 85.0\% of the cases below 2$^{24}$, as Table \ref{tab:finalcomp} shows.  What is less clear, given the scale of the axes, is that \textit{all} of the solutions are now unique.  This is due to the more complex prefactors involved.

\begin{table}
\begin{ruledtabular}
\begin{tabular}{c|cccc}
$q$ & 2 & 3 & 4 & 5 \\
\hline
Optimal & 1,687 & 14,253,472 & 2,522,055 & 1\\
Unique & 1,687 & 14,253,472 & 2,522,055 & 1\\
\end{tabular}
\caption{Performance of various $q$-sectioning search heuristics for 1/$\epsilon \in \left\{2, 3, ..., 2^{24}\right\}$ and $\alpha_{0}=0.99$. As per Table \ref{tab:naive}, performance is optimal where no $q$-sectioning heuristic solution exists for that $\epsilon^{-1}$ with fewer measurements, and unique if no similar solution for that $\epsilon^{-1}$ exists with different $q$.}
\label{tab:finalcomp}
\end{ruledtabular}
\end{table}

It is now trivial to compare the estimates of the means from Table \ref{tab:more} with Eq. \ref{eq:behaviour} -- see Table \ref{tab:meancomp}.  The agreement is very good, with most estimates within 1 measurement of the corresponding analytic values.

\begin{table}
\begin{ruledtabular}
\begin{tabular}{cc|cccc}
$\epsilon^{-1}$ & Type & \multicolumn{4}{c}{$q$}\\
\multicolumn{2}{c|}{} & 2 & 3 & 4 & 5 \\
\hline
\multirow{2}{*}{$10^{3}$} & Estimate & 1994.7 & 2092.7 & 1992.0* & 2489.5\\
& Analytic & 1995 & 2093 & 1992.5* & 2490\\
\rule{0pt}{2.6ex}
\multirow{2}{*}{$3\times 10^{3}$} & Estimate & 2394.6 & 2391.7 & 2389.7* & 2429.5\\
& Analytic & 2394 & 2392 & 2391* & 2490\\
\rule{0pt}{2.6ex}
\multirow{2}{*}{$10^{4}$} & Estimate & 2797.7 & 2690.3* & 2787.8 & 2989.1\\
& Analytic & 2793 & 2691* & 2789.5 & 2988\\
\rule{0pt}{2.6ex}
\multirow{2}{*}{$3\times 10^{4}$} & Estimate & 2992.8 & 2988.7* & 3188.9 & 3845.9\\
& Analytic & 2992.5 & 2990* & 3188 & 3486\\
\rule{0pt}{2.6ex}
\multirow{2}{*}{$10^{5}$} & Estimate & 3391.9 & 3288.8* & 3584.9 & 3983.7\\
& Analytic & 3391.5 & 3289* & 3586.5 & 3984\\
\rule{0pt}{2.6ex}
\multirow{2}{*}{$3\times 10^{5}$} & Estimate & 3790.7 & 3587.0* & 3985.2 & 3984.9\\
& Analytic & 3790.5 & 3588* & 3985 & 3984\\
\rule{0pt}{2.6ex}
\multirow{2}{*}{$10^{6}$} & Estimate & 3990.2 & 3887.4* & 3983.5 & 4482.8\\
& Analytic & 3990 & 3887* & 3985 & 4482\\\end{tabular}
\caption{Comparison of means, estimated from 10$^{6}$ Monte Carlo trials, versus those calculated directly from Eq. \ref{eq:behaviour}.  The lowest values in each column are denoted with asterisks (*).}
\label{tab:meancomp}
\end{ruledtabular}
\end{table}


\subsection{Behaviour with changes in loss}
\label{sec:loss}

\begin{figure*}[tb!]
\centering
\includegraphics[angle=0, width=0.99\linewidth]{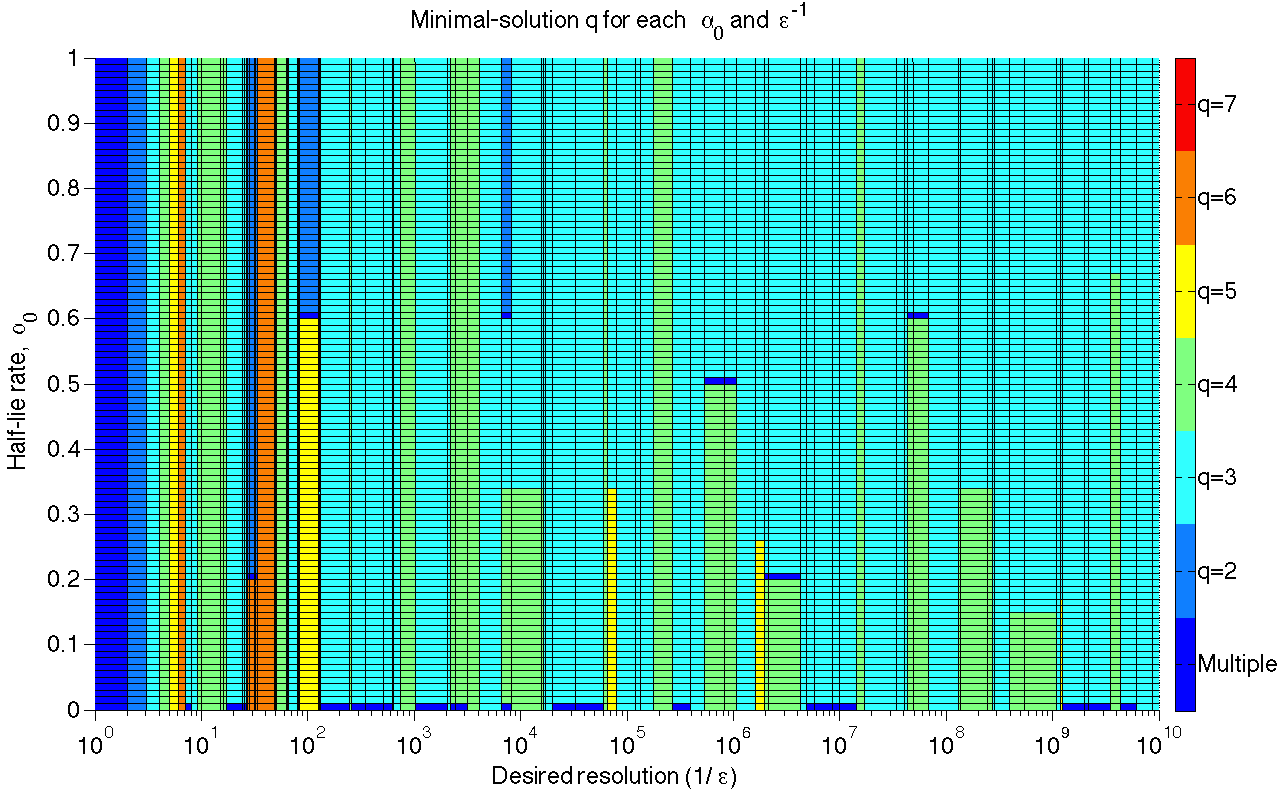}
\caption{Minimal $q$-sectioning with verification heuristic; $q$:Min$_{q}\left(\bar{m}_{q,\epsilon^{-1},\alpha_{0}}\right)$ for each $\epsilon^{-1}\in\left\{2,3,...10^{10}\right\}$ and\qquad\qquad\qquad\qquad $\alpha_{0}\in\left\{0,0.01,...,0.99\right\}$.  The $\epsilon^{-1}$ grid is set at a unique sorted list of $\epsilon^{-1}\in\left\{q^{i}-1, q^{i}, q^{i}+1\right\}$ for $q\in\left\{2,3,4,5\right\}$ and for any $i$ giving a value of $\epsilon^{-1}$ in the desired range.   Panel colours show the minimal solution type; panels are shaded by the solution type(s) at their lower-left vertices, although due to the step-like nature of $\bar{m}_{q,\epsilon^{-1},\alpha_{0}}$ the solution types are valid for all $\epsilon^{-1}$ across each panel horizontally.  Note: no solutions are given for $\alpha_{0}=1$ as these are undefined.}
\label{fig:general}
\end{figure*}

Obtaining a general analytic form allows us to explore behaviour of the family of $q$-sectioning with verification heuristics across changes not only to the desired precision, but also of the half-lie rate $\alpha_{0}$.  Due to the step-like nature of the $\bar{m}_{q,\epsilon^{-1},\alpha_{0}}$ function, its behaviour can be modelled simply by taking values on each side of the step function locations -- which are the powers of the various $q$.  Figure \ref{fig:general} shows the optimal heuristic type(s) for each $\epsilon^{-1}\in\left\{2,3,...10^{10}\right\}$ and $\alpha_{0}\in\left\{0,0.01,...,1\right\}$.

The abundance of light blue in Fig. \ref{fig:general} shows that the trisectioning with verification heuristic is optimal not only for most $\epsilon$ at $\alpha_{0}=0.99$, but also at lower half-lie rates.  This is a somewhat surprising result; above $\epsilon^{-1}=128$, trisectioning dominates the plot - except for zero half-lies.  Of course, the $q$-sectioning with verification heuristic is not optimal at zero half-lies, since it does not take advantage of the information contained in failures to detect photons; basic bisectioning (binary search) search is faster there.  However, searches not requiring verification cannot be 100\% fidelitous for any $\alpha_{0}>0$.

As we might expect, the next most successful heuristic is tetrasectioning with verification.  Bisectioning makes a few brief appearances, but is not uniquely optimal for any $\alpha_{0}$ after $\epsilon=2^{13}$.  Pentasectioning is similar, though rarer again, and is no longer uniquely optimal under any conditions after $q=5^{13}$.

In principle, Fig. \ref{fig:general} could be extended further in $\epsilon^{-1}$; however near $2^{49}$, double-precision floats fail to avoid numerical errors in calculating $\bar{m}_{q,\epsilon^{-1},\alpha_{0}}$.  Extending Fig. \ref{fig:general} this far would unacceptably compress the plot, which for clarity is limited to cases already discussed.


\section{Conclusions}
\label{sec:conc}

In summary, the $q$-sectioning with verification family of search algorithms has been proposed and studied.  Verification is an intriguing constraint on search algorithms; its cost is most apparent at low photon loss rates.  For zero loss, bisectioning with verification ignores half of the available information and suffers a consequent loss of efficiency compared to the optimal binary search.  However, for any non-zero loss, algorithms requiring verification retain 100\% fidelity in their predictions, while $q$-ary searches cannot (except in the unattainable limit of infinite measurements).

An analytic formula describing the mean number of measurements for entire the family of $q$-sectioning with verification algorithms given any choice of $q$, precision $\epsilon$, and half-lie (photon loss) rate $\alpha_{0}$ has been developed.  The family's behaviour has been mapped across a wide range of $\epsilon$ and $\alpha_{0}$, and the optimal choice is not simply a function of the relative nearness of $\epsilon$ to a power of $q$.

Trisectioning with verification is most often the optimal $q$-sectioning with verification heuristic for problems involving any random half-lies and precision $\epsilon<1/128$ of the initial search domain.  Further, although trisectioning with verification is sub-optimal, it appears near-optimal for our example case according to our estimator.

In practical terms, since photon-counting microscopes have already been developed, the trisectioning scheme discussed here offers the chance for greatly accelerated target searches over current confocal rastering techniques.  This approach could be particularly useful for locating short-lived species if the microscope point-spread function can be modified quickly enough, via both changing the spot size as well as shifting the beam centre, thereby changing the search domain rapidly between measurements.  However, before such a task is undertaken, it would be beneficial to reconsider this problem with explicit Gaussian roll-offs and/or general shape to the PSF, and also to fully characterise the intensity changes required by the varying PSF widths.  Finally, before such an approach is applied to biological systems, attempts should be made to include penalty functions characterising photon exposure and/or damage, as well as accounting for practical considerations such as time to modify or shift the PSF, to study how the dynamics of an optimal biological search differ from those presented here.  Information may not be the sole metric of interest to be optimised; a full cost-benefit analysis accounting for these issues may eschew the information benefit of shifting position more often.

\section*{Acknowledgements}

The authors thank Nicolas Menicucci, Brant Gibson, Adrian Dyer, Jair Garcia, and Chris Xu for insightful discussions, and acknowledge the support of the Australian Research Council (project numbers CE140100003 and FT160100357).

\section*{Author Contributions}
DWD and ADG conceived of, planned, carried out, and analysed the study.  DWD and ADG wrote the manuscript.\\
\\
\section*{Additional Information}
The authors declare no competing financial interest.


\begin{thebibliography}{100}

\bibitem{Pawley06} J.~B.~Pawley, \textit{Handbook of Biological Confocal Microscopy}, Springer, 2006.

\bibitem{Horton13} N.~G.~Horton, K.~Wang, D.~Kobat, C.~G.~Clark, F.~W.~Wise, C.~B.~Schaffer, and C. Xu, ``\textit{In vivo} three-photon microscopy of subcortical structures within an intact mouse brain'', \textit{Nature Photonics}, {\bf 7}:205, 2013.

\bibitem{Chalfie94} M.~Chalfie, Y.~Tu, G.~Euskirchen, W.~W.~Ward, and D.~C.~Prasher, ``Green Fluorescent Protein as a Marker for Gene Expression'', \textit{Science} {\bf 263}:802, 1994.

\bibitem{Reineck16} P.~Reineck and B.~C.~Gibson, ``Near-Infrared Fluorescent Nanomaterials for Bioimaging and Sensing'', \textit{Advanced Optical Materials}, 2016.

\bibitem{Reineck16a} P.~Reineck, A.~Francis, A.~Orth, D.~W.~M.~Lau, R.~D.~V.~Nixon-Luke, I.~Das~Rastogi, W.~A.~W.~Razali, N.~M.~Cordina, L.~M.~Parker, V.~K.~A.~Sreenivasan, L.~J.~Brown, and B.~C.~Gibson, ``Brightness and Photostability of Emerging Red and Near-IR Fluorescent Nanomaterials for Bioimaging'', \textit{Advanced Optical Materials} {\bf 4}(10):1549, 2016.

\bibitem{Kochevar81} I.~E.~Kochevar, ``Phototoxicity Mechanisms: Chlorpromazine Photosensitized Damage to DNA and Cell Membranes'', \textit{The Journal of Investigative Dermatology}, {\bf 76}:59, 1981.

\bibitem{Song95} L.~Song, E.~J.~Hennink, I.~T.~Young, and H.~J.~Tanke, ``Photobleaching Kinetics of Fluorescein in Quantitative Fluorescence Microscopy'', \textit{Biophysical Journal}, {\bf 68}:2588, 1995.

\bibitem{Shaner08} N.~C.~Shaner, M.~Z.~Lin, M.~R.~McKeown, P.~A.~Steinbach, K.~L.~Hazelwood, M.~W.~Davidson, and R.~Y.~Tsien, ``Improving the photostability of bright monomeric orange and red fluorescent proteins'', \textit{Nature Methods}, {\bf 5}(6):545, 2008.

\bibitem{Nowak08} R.~Nowak, ``Generalized binary search'', \textit{Communication, Control, and Computing, 2008 46th Annual Allerton Conference on.}, IEEE, 2008.

\bibitem{Nowak09} R.~Nowak, ``Noisy generalized binary search'', in \textit{Advances in neural information processing systems}, p1366, Curran Associates, Inc., 2009.

\bibitem{Garey74} M.~R.~Garey and R.~L.~Graham, ``Performance Bounds on the Splitting Algorithm for Binary Testing'', \textit{Acta Informatica}, {\bf 3}:347, 1974.

\bibitem{Renyi61} A.~R\'enyi, ``On a problem of information theory'', \textit{MTA Mat. Kut. Int. Kozl.} {\bf 6B}:505, 1961.

\bibitem{Ulam76} S.~M.~Ulam, \textit{Adventures of a Mathematician}, p. 281, \textit{Scribner}, New York, 1976.

\bibitem{Pelc02} A.~Pelc, ``Searching games with errors -- fifty years of coping with liars'', \textit{Theoretical Computer Science}, {\bf 270}:71, 2002.

\bibitem{Ellis04} R.~Ellis and C.~Yan, ``Ulam's pathological liar game with one half-lie'', \textit{Int. J. Math. Math. Sci.}, {\bf 2004}(09):1523, 2004.

\bibitem{Ellis05} R.~B.~Ellis, V.~Ponomarenko, and C.~H.~Yan, ``The R\'enyi-Ulam pathological liar game with a fixed number of lies'', \textit{J. Comb. Th. A}, {\bf 112}:328, 2005.

\bibitem{Ellis08} R.~B.~Ellis, V.~Ponomarenko, and C.~H.~Yan, ``How to play the one-lie R\'enyi-Ulam game'', \textit{Disc. Math.}, {\bf 308}:5805, 2008.

\bibitem{Cicalese13} F.~Cicalese, \textit{Fault-Tolerant Search Algorithms: Reliable Computation with Unreliable Information}, Springer-Verlag, 2013.

\bibitem{Yao85} A.~C.~Yao and F.~F.~Yao, ``On fault-tolerant networks for sorting'', \textit{SIAM J. Comput.}, {\bf 14}:120, 1985.

\bibitem{Ravikumar87} B.~Ravikumar, K.~Ganesan, and K.~B.~Lakshmanan, ``On Selecting the Largest Element In Spite of Erroneous Information'', \textit{Annual Symposium on Theoretical Aspects of Computer Science}, Springer Berlin Heidelberg, 1987.

\bibitem{Lakshmanan91} K.~B.~Lakshmanan, B.~Ravikumar, and K.~Ganesan, ``Coping with erroneous information while sorting'', \textit{IEEE Trans. Comput.}, {\bf 40}:1081, 1991.

\bibitem{Feige94} U.~Feige, D.~Peleg, P.~Raghavan, and E.~Upfal, ``Computing with noisy information'', \textit{SIAM J. Comput.}, \textit{23}:1001, 1994.

\bibitem{DeBonis97} A.~De~Bonis, L.~Gargano, and U.~Vaccaro, ``Group testing with unreliable tests'', \textit{Inform. Sci.}, {\bf 96}:1, 1997.

\bibitem{Ngo00} H.~Q.~Ngo and D.~Z.~Du, ``A Survey on Combinatorial Group Testing Algorithms with Applications to DNA Library Screening'', in \textit{Discrete Mathematical Problems with Medical Applications: DIMACS Workshop}, {\bf 55}:171, Am. Math. Soc., 2000.

\bibitem{Cicalese03} F.~Cicalese and D.~Mundici, ``Learning and the Art of Fault-Tolerant Guesswork'', in \textit{Adaptivity and Learning: An Interdisciplinary Debate'}, Eds. R.~K\"uhn, R.~Menzel, W.~Menzel, U.~Ratsch, M.~M.~Richter, and I.-O.~Stamatescu, Springer-Verlag, 2003.

\bibitem{Mancini05} S.~Mancini and L.~Maccone, ``Using Quantum Mechanics to Cope with Liars'', \textit{Int. J. Quant. Inf.}, {\bf 3}(4):729, 2005.

\bibitem{Karp07} R.~M.~Karp and R.~Kleinberg, ``Noisy binary search and its applications'', in \textit{Proc. 18$^{{\rm th}}$ ACM-SIAM symposium on discrete algorithms}, Soc. Ind. and Appl. Math., 2007.

\bibitem{Corsi16} E.~A.~Corsi and F.~Montagna, ``The R\'enyi--Ulam games and many-valued logics'', \textit{Fuzzy Sets and Systems}, {\bf 301}:37, 2016.

\bibitem{Jedynak11} B.~Jedynak, P.~I.~Frazier, and R.~Sznitman, ``Twenty questions with noise: Bayes optimal policies for entropy loss'', \textit{J. Appl. Prob.}, {\bf 49}(1):114, 2011.

\bibitem{Pelc89} A.~Pelc, ``Searching with known error probability'', \textit{Th. Comp. Sci.}, {\bf 63}:185, 1989.

\bibitem{Rivest80} R.~L.~Rivest, A.~R.~Meyer, D.~J.~Kleitman, and K.~Winklmann, ``Coping with Errors in Binary Search Procedures'', \textit{J. Comp. Sys. Sci.}, {\bf 20}:396, 1980. 

\bibitem{Ravikumar84} B.~Ravikumar and K.~B.~Lakshmanan. ``Coping with known patterns of lies in a search game'', \textit{Theoretical Computer Science} {\bf 33}(1):85, 1984.

\bibitem{Pelc87} A.~Pelc, ``Solution of Ulam's Problem on Searching with a Lie'', \textit{Journal of Computational Theory Series A}, {\bf 44}:129, 1987.

\bibitem{Muthukrishnan94} S.~Muthukrishnan, ``On optimal strategies for searching in the presence of errors'', in \textit{Proceedings of the Fifth ACM-SIAM symposium on Discrete algorithms}, p680, 1994.

\bibitem{Dhagat92} A.~Dhagat, P.~G\'acs, and P.~Winkler, ``On playing ``twenty questions'' with a liar'', in \textit{Proc. 3$^{{\rm rd}}$ ACM-SIAM symposium on discrete algorithms}, Soc. Ind. Appl. Math., 1992.

\bibitem{Cicalese00} F.~Cicalese and D.~Mundici, ``Optimal Coding with One Asymmetric Error: Below the Sphere Packing Bound'', in \textit{Computing and Combinatorics: Proc. 6$^{{\rm th}}$ Int. Comp. Comb. Conf.}, Eds. D.-Z.~Du, P.~Eades, V.~Estivill-Castro, X.~Lin, and A.~Sharma, Springer-Verlag, 2000.

\bibitem{Cicalese03a} F.~Cicalese and C.~Deppe, ``Quasi-Perfect Minimally Adaptive q-ary Search with Unreliable Tests'', in \textit{Proceedings of the 14$^{th}$ International Symposium on  Algorithms and Computation}, Springer-Verlag, 2003.

\bibitem{Cicalese04} F.~Cicalese, C.~Deppe, and D.~Mundici, ``Q-Ary Ulam-R\'enyi Game with Weighted Constrained Lies'', in \textit{Proceedings of the 10$^{th}$ International Computing and Combinatorics Conference}, Springer-Verlag Berlin, 2004. 

\bibitem{Cicalese07} F.~Cicalese and C.~Deppe, ``Perfect minimally adaptive q-ary search with unreliable tests'', \textit{Journal of Statstical Planning and Inference} {\bf 137}:162, 2007.

\bibitem{Xing16} S.~M.~Xing, W.~A.~Liu, and K.~Meng, ``R\'enyi-Berlekamp-Ulam searching game with bi-interval queries and two lies'', \textit{Discrete Applied Mathematics} {\bf 202}:8, 2016.

\bibitem{Schalkwijk71} J.~P.~M.~Schalkwijk, ``A Class of Simple and Optimal Strategies for Block Coding on the Binary Symmetric Channel with Noiseless Feedback'', \textit{IEEE Transactions on Information Theory}, {\bf IT-17}(3):283, 1971.

\bibitem{Peterson54} W.~W.~Peterson, T.~G.~Birdsall, and W.~C.~Fox, ``The Theory of Signal [\textit{sic}] Dectectability'', \textit{Transactions of the IRE professional group on information theory}, {\bf 4}(4):171, 1954

\bibitem{Green66} D.~M.~Green and J.~A.~Swets, \textit{Signal detection theory and psychophysics}, \textit{Wiley}, New York, 1966.

\bibitem{Spencer92} J.~Spencer and P.~Winkler, ``Three thresholds for a liar'', \textit{Combinatorics, Probability and Computing} {\bf 1}:81, 1992.

\bibitem{Soskov89} I.~N.~Soskov, ``Definability via Enumerations'', \textit{J. Symb. Logic}, {\bf 54}(2):428, 1989.

\bibitem{Dobkin91} D.~P.~Dobkin and D.~L.~Souvaine, ``Detecting the intersection of convex objects in the plane'', \textit{Computer aided geometric design}, {\bf 8}(3):181, 1991.

\bibitem{Salehin10} K.~M.~Salehin and R.~Rojas-Cessa, ``Combined methodology for measurement of available bandwidth and link capacity in wired packet networks'' \textit{IET communications}, {\bf 4}(2):240, 2010.

\bibitem{Bentley97} J.~L.~Bentley and R.~Sedgewick, ``Fast Algorithms for Sorting and Searching Strings'', \textit{SODA}, {\bf 97}:360, 1997.

\end{thebibliography}
\end{document}